# Moore's law, Wright's law and the Countdown to Exponential Space

By Daniel Berleant, Venkat Kodali, Richard Segall, Hyacinthe Aboudja, and Michael Howell

Technologies have often been observed to improve exponentially over time (Nagy et al. 2013). In practice this often means identifying a constant known as the doubling time, describing the time period over which the technology roughly doubles in some measure of performance or of performance per dollar. Moore's law is, classically, the empirical observation that the number of electronic components that can be put on a chip doubles every 18 months to 2 years (Moore 1965). Today it is frequently stated as the number of computations available per unit of cost [1]. Generalized to the appropriate doubling time, it describes the rate of advancement in many technologies. A frequently noted competitor to Moore's law is known as Wright's law, which has aeronautical roots (Wright 1936). Wright's law (also called power law, experience curve and Henderson's law) relates some quality of a manufactured unit (for Wright, airplanes) to the volume of units manufactured. The equation [8] expresses the idea that performance — price or a quality metric — improves according to a power of the number produced, or alternatively stated, improves by a constant percentage for every doubling of the total number produced.

Does exploration of outer space conform to Moore's law or Wright's law-like behavior? Our results below are broadly consistent with these laws. This is true for many technologies (Nagy 2013). Although the two laws can make somewhat different predictions, Sahal (1979) found that they converge to the same predictions when manufacturing volume increases exponentially over time. When space exploration transitions into an independent commercial sector, as many people hope and expect, spacecraft technology will then likely enter an era of unambiguously exponential advancement.

The concept of exponential increases in space exploration has a checkered history. On the one hand, the lure of space and the impatience of so many to see humankind expand faster into that wide open frontier creates the desire for exponential space. On the other hand, progress may at times seem painfully slow. Turner (2004) pleaded for "Permission to believe" that space launches would follow a Moore's law (i.e., increase exponentially over time). On the question of crewed flights in particular, Hicks (2015) states it is "... sad that human space exploration has stalled." Bardi (2016) shows a startlingly pessimistic graph of crewed flights, stating that "the data are clear and cannot be ignored: human spaceflight is winding down." Yet there is room for argument. Analyses by SpacePunx (2016) and Elliott (2014) suggest a generally upward, if not necessarily exponential, trendline in the human population of space, but Flo422 (2018) gives data more supportive of Bardi's negative outlook. A sampling of further broadly pessimistic commentary on the rate of advancement in space exploration appears in [2].

Roberts (2011) attributes much of the pessimism to reliance on the wrong proxies for progress, arguing that advancement is actually proceeding apace. Clearly "advancement in space exploration" is a concept, not a directly measurable primary or derived variable like distance, cost, distance per unit cost, launch year, and so on. Thus, the appropriate choice of a quantifiable proxy variable [3] is a critical issue (Turchin 2018). A poor proxy might be simply irrelevant (e.g. reflectance of the spacecraft exterior). It might show little or no change over time (launch mass of extraterrestrial mission spacecraft [4]). Or it might show (Rupp 2018) a trend that ends too soon (for microprocessors, clock frequency) even though other proxies show a continuing trend, or starts trending with too long a delay (microprocessor core count).

Our search for proxies has shown that multi-parameter derived variables, despite the risk of overfitting (Nagy et al. 2013), show promise (Hall et al. 2017). Our strategy here, which avoids the overfitting issue, is to identify a simple primary variable that shows a useful trend. The focus here is on one such variable: *spacecraft lifespan*.

Initially it would be natural to suggest graphing spacecraft lifespan vs. launch year to check if lifespan is increasing over time. However, this metric has an unfixable bias problem. Any year which launched at least one spacecraft that is still operational provides lifespan data only for the shorter-lived craft that are no longer working. This underestimates the lifespans that that year's technology was able to produce. The problem occurs starting as early as 1977 when Voyager 1 and 2 were launched. These craft are still operational and transmitting data (indeed, as of Nov. 5, 2018, Voyager 2, the slower one, crossed the heliopause putting both officially in interstellar space [5]).

The solution discussed below is to shift from associating spacecraft lifespans and their launch dates to, instead, associating lifespans with the dates they ended operation. Graphing the resulting data points, we hypothesized and our results show, could reveal a trend of increasing lifespans over time. The data points used were spacecraft whose activies included exploration of at least one extraterrestrial body (except the Sun; those craft may be incorporated into a revised analysis in the future). A spacecraft was defined prior to the analysis to include those modules that were physically connected at their beginning of operational life. Thus, two separate spacecraft could be launched by the same launch vehicle. One example is GRAIL A and GRAIL B (Gravity Recovery and Interior Laboratory A and B), moon missions launched together on Sept. 10, 2011. Another is the Oct. 23, 2014 co-launched Manfred Memorial Moon Mission and Chang'e 5-T1. On the other hand, multiple physically separating objects could count as a single spacecraft. An example is the Mars Exploration Rover A (MER-A), or Spirit, which contained a cruise stage, lander, and rover. MER-B, or Opportunity, had the same design. A more extreme example is asteroid explorer Hayabusa2, a mothership containing four rovers carried by two rover deployment landers (MINERVA-II-A and MINERVA-II-B), the Small Carry-On Impactor, a smaller tantalum impactor, and a sample return capsule. It will also eject multiple reflective target markers to aid in landing the mothership and leave behind the DCAM3 camera. These spacecraft and the others contributing to the analyses are in a publicly accessible spreadsheet [6].

**Moore's law and Wright's law for extraterrestrial exploration spacecraft**

To test how well the data conforms to a Moore's law trend for spacecraft lifespan, we graphed the lifetimes in years of spacecraft exploring extraterrestrial bodies (except the Sun) vs. the time point marking their end. See Fig. 1. The regression line implies a doubling time of 7 years for spacecraft lifespan.

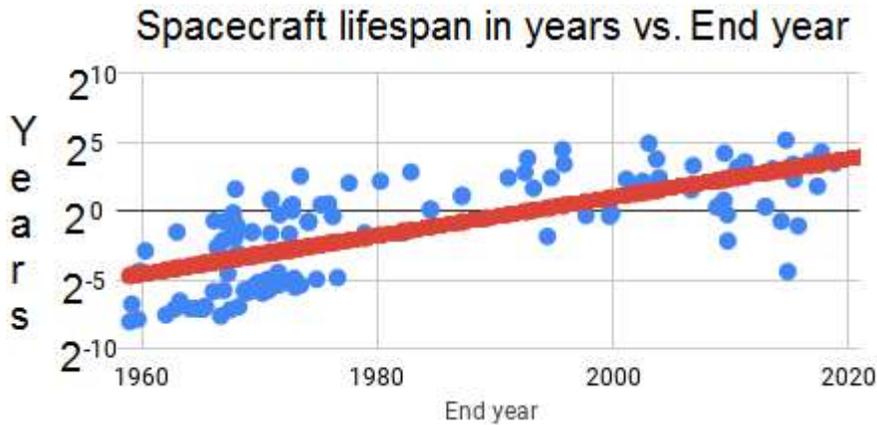

**Figure 1**. Lifespans of extraterrestrial exploration spacecraft. Each craft's lifespan end date is plotted as a year and fraction thereof. Each lifespan is plotted as number of years, so $2^0$=1 year, $2^5$=32 years, $2^{-5}$=1/32 year=11.41 days, etc. The Moore's law regression line is:
**Lifespan in years = $2^{-4.6642+0.13817*(End\ year\ -\ 1959)}$**.

Inspection of Fig. 1 suggests a dramatic improvement in lifespan until the 1990s followed by little further improvement. Possible reasons for this include:
1. Newly spacefaring organizations might be expected to start with spacecraft with shorter lifespans, lowering the average for the more recent region of the graph. For example the outlier data point in 2014 with an low lifespan of approximately $2^{-5}$ years was the Manfred Memorial Moon Mission (M4), a project of the Luxembourg-based company LuxSpace and the first commercial extraterrestrial spaceflight. That's an advance, even if on the graph it looks more like a step back.
2. Lifespan improvement may have stalled.

Fig. 2 suggests both may play a factor, as a graph of just NASA-led missions indicates NASA's lifespan increases have slowed but not stopped, with the red regression line for the post-1990 period defining a doubling time 17 years for spacecraft lifespan.

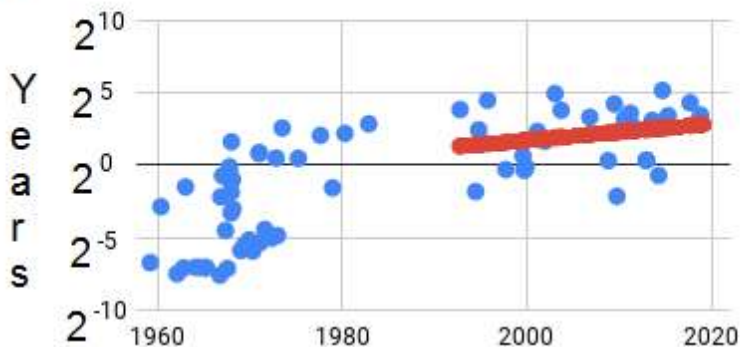

**Figure 2**. NASA-led spacecraft lifespans with Moore's law regression for 1992-present. A gap from 1983 to 1991 is associated with a NASA emphasis on non-extraterrestrial missions, including the Space Shuttle program. The Moore's law regression equation is:
**Lifespan in years = $2^{1.154+0.0576*(End\ year\ -\ 1990))}$**.

What about Wright's law? Fig. 3 tells the story so far. Wright's law is based on production volume, but modeling the first spacecraft to explore an extraterrestrial object (Luna 1, destination the Moon) as representing a volume of 1 has poor validity, because Luna 1 benefited from the world's experience producing seven prior craft which were failures, thus did not explore an extraterrestrial object, hence are not otherwise used in our analysis. (These failed missions were Pioneer 0, 1, 2, and 3, and Luna E1 #1, #2, and #3.) Luna 1, performing the first extraterrestrial exploration, was thus modeled as representing a volume of 8, with each subsequent mission adding one to the overall volume produced of extraterrestrial exploration spacecraft. The exponent of the regression curve in Fig. 3 is 2.528, which implies lifespan increasing by a factor of 6 for each doubling of production volume. Equivalently, lifespan doubles with each increase in volume of 32%. Our modeling did not account for failures after the first seven, although the early years of the US/USSR space race were marked by a lot of them.

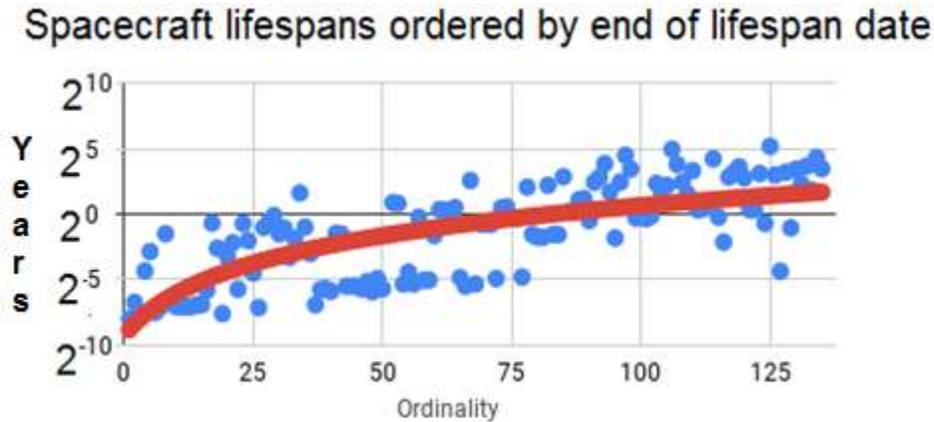

**Figure 3**. Ordinality is the number of the spacecraft, ordered by end date of the spacecraft's lifespan. The red regression curve fits the Wright's law formula to the data. The Wright's law equation is: Lifespan in years = $1.143 \times 10^{-5} \times (\text{Ordinality}+7)^{2.528}$.

Which model, Moore or Wright, fits the data better? Comparing data points to regression curves, the RMS error in Fig. 1 (Moore) was 2.485, while the RMS error in Fig. 3 (Wright) was slightly but perhaps not significantly better at 2.459. This is consistent with the result of Nagy et al. (2013) who, analyzing many varied types of technology, found that Moore and Wright both work close to equally well but with Wright having a slight edge.

**From extraterrestrial spacecraft to satellites**

A useful question is whether results obtained for one kind of space endeavor might apply to other kinds. Investigating this, we examined comprehensive satellite data (McDowell). Many, many more satellites have been launched than extraterrestrial exploration spacecraft. From the data field "Current Status" for each satellite, we excluded the 42.68% of satellites with status "In Earth orbit" the vast majority of which leave the "Date of Status" field blank. For the remaining 57.32% of the satellites (most with status "Reentered" at 51.57% of all satellites), the Date of Status field then gives a year which we used as a proxy for the end of its lifespan. The difference between that year and the year in the Launch Date field was used to estimate the lifespan. The analysis resulted in Fig. 4.

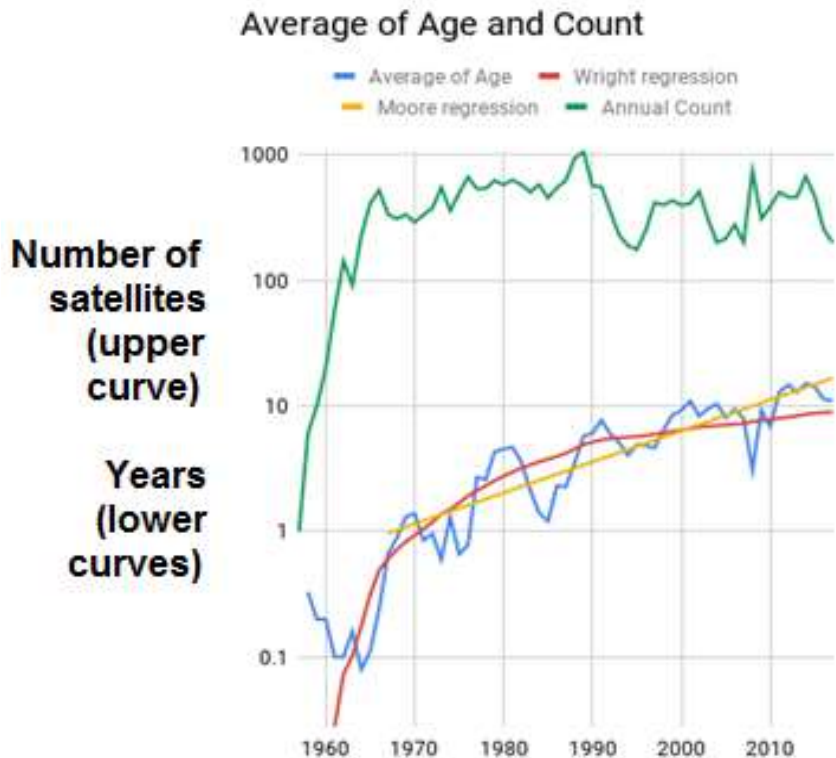

**Figure 4**. Mean lifespan (blue curve) of satellites whose lifespan ended in a given year. The number of satellites ending in each year (green curve) rose rapidly then plateaued. A Moore's law curve (in yellow, Lifespan in years = $0.549*2^{(End\ year-1957)/12.17}$) was fitted to the data. A Wright's law curve, in red, with equation Lifespan in years = $0.0002446*ordinality^{1.04}$, where the ordinality associated with each year is determined from the number of satellite lifespans ending in that year (green curve), was also fitted. The source spreadsheet is available [7].

In Fig. 4, note the disparity in the future trends suggested by the Moore and Wright curves: upward, but by how much? The red Wright's law curve has some discernible bumpiness associated with obvious bumps in the green curve. It begins in 1961 because data for earlier years suffered from small numbers of satellites and thus the risk of noisy data, short lifespans and thus less reliable lifespan calculation due to the graininess of looking at launch and end year but not month or day, and initialization effects, all leading to a curve fitting process that would result in a relatively poor fit to later data, although that fit to later data is of primary interest from the standpoint of predicting future satellite lifespans. There were 58 satellites contributing to the 1961 data and more for later years, while 1957 provided only one satellite, 1958 six, 1959 ten, and 1960 twenty. The Moore curve begins in 1967 due to initialization effects that tend to distort the starts of exponential trends in general and, here, would give the disparity between the curve and earlier years' data an inordinate effect on the RMS error minimization process, yielding a poorer fit to the later years' data and thus poorer insight into future years.

The Moore's law curve implies that lifespan has a doubling time of 15 years. Compare that to the extraterrestrial spacecraft lifespan doubling time (based on Fig. 1) of 7 years overall, and 17 years (based on Fig. 2) for post-1990 NASA mission craft. For Wright's law, the data implies approximate doubling of satellite lifespan for every doubling in cumulative satellite production (more precisely, in cumulative qualifying satellite lifespan end data).

It is important to notice that the satellite analysis excludes satellites whose lifespans ended but are still in orbit, because the source data, while it states certain status changes in the satellites, does not directly state when its useful life is over. Some satellites, becoming inert lumps of matter, remain in orbit. This provides a potentially huge bias in the data. The fact that both Moore and Wright curves in Fig. 4 yield interesting fits could be interpreted to mean not only that satellite lifespan is increasing over time, but also that planet4589 Change of Status data are useful for analyzing satellite lifespan.

**Conclusions**

Spacecraft lifespan is a useful proxy for advancement in space exploration technology. It permits fitting of Moore's law (exponential) and Wright's law (power) curves to the data that show dramatic improvement over the course of the spacefaring era. The dream that satellites might last 100 years (Gonzalo et al. 2014) will come true if (again, if) present trends continue. For extraterrestrial exploration craft, one path to dramatically increasing spacecraft lifespans would be to construct human colonies on the Moon or even Mars that last indefinitely and use the craft as part of the physical structure of the colony. Another would be reusable craft that take passengers safely to the Moon and back many times. Still another would be commercialization (e.g. tourism, mining operations, etc.) resulting in large increases in spacecraft manufacturing volume and consequent gains in technological performance based on Wright's law. The historical trends bode well for significant future improvement in spacecraft lifespan but the detailed technologies have yet to be innovated.

**About the Authors**

Daniel Berleant, PhD (berleant@gmail.com), is Professor of Information Science, University of Arkansas at Little Rock. Venkat Kodali, MS, is Director of Business Intelligence at CARTI in Little Rock. Richard Segall, PhD, is Professor of Computer and Information Technology at Arkansas State University. Hyacinthe Aboudja, PhD, is Assistant Professor of Computer Science at Oklahoma City University. Michael Howell is an Information Science master's candidate, University of Arkansas at Little Rock.